\pdfoutput=1

\documentclass[journal]{IEEEtran}
\ifCLASSOPTIONcompsoc
  \usepackage[nocompress]{cite}
\else
  \usepackage{cite}
\fi
\ifCLASSINFOpdf
  \usepackage[pdftex]{graphicx}
\else
\fi
\usepackage{booktabs}
\usepackage{array}
\usepackage{framed,multirow}
\usepackage{multicol}
\usepackage{amssymb}
\usepackage{amsmath}
\usepackage{latexsym}
\usepackage{url}
\usepackage{xcolor}
\usepackage{hyperref}
\usepackage{todonotes}

\newcommand{\eg}{{e.g.}}           
\newcommand{\ie}{{i.e.}}           

\begin{document}
%
\title{MetricUNet: Synergistic Image- and Voxel-Level Learning for Precise Prostate Segmentation via Online Sampling}
%
%
%
%

\author{Kelei~He,~Chunfeng~Lian,~Ehsan~Adeli,~Jing~Huo,~Yang~Gao,~Bing~Zhang,\\Junfeng~Zhang*,~Dinggang~Shen*,~\IEEEmembership{Fellow,~IEEE}
\IEEEcompsocitemizethanks{\IEEEcompsocthanksitem K. He and J. Zhang are with Medical School of Nanjing University, Nanjing, P. R. China. E. Adeli is with the Department of Psychiatry and Behavioral Sciences and the Department of Computer Science at Stanford University, CA, USA. J. Huo, Y. Gao are with the State Key Laboratory for Novel Software Technology, Nanjing University, Nanjing, P. R. China. K. He, Y. Gao and J. Zhang are also with the National Institute of Healthcare Data Science at Nanjing University, Nanjing, P. R. China. B. Zhang is with Department of Radiology, Nanjing Drum Tower Hospital, Nanjing University Medical School, P. R. China. C. Lian is with School of Mathematics and Statistics, Xi'an Jiaotong University, Shanxi, China. D. Shen is with School of Biomedical Engineering, ShanghaiTech University, Shanghai, China. He is also with the Department of Research and Development, Shanghai United Imaging Intelligence Co., Ltd., Shanghai, China. He is also with the Department of Artificial Intelligence, Korea University, Seoul, Republic of Korea.}
\thanks{* Corresponding authors: Junfeng Zhang (jfzhang@nju.edu.cn); Dinggang Shen (dgshen@med.unc.edu)}
}

%
%

\markboth{Journal of \LaTeX\ Class Files,~Vol.~14, No.~8, August~2015}%
{Shell \MakeLowercase{\textit{et al.}}: Bare Demo of IEEEtran.cls for Computer Society Journals}
%



\IEEEtitleabstractindextext{%
\begin{abstract}
Fully convolutional networks (FCNs), including UNet and VNet, are widely-used network architectures for semantic segmentation in recent studies. However, conventional FCN is typically trained by the cross-entropy or Dice loss, which only calculates the error between predictions and ground-truth labels for pixels individually. This often results in non-smooth neighborhoods in the predicted segmentation. This problem becomes more serious in CT prostate segmentation as CT images are usually of low tissue contrast. To address this problem, we propose a two-stage framework, with the first stage to quickly localize the prostate region, and the second stage to precisely segment the prostate by a multi-task UNet architecture. We introduce a novel online metric learning module through voxel-wise sampling in the multi-task network. Therefore, the proposed network has a dual-branch architecture that tackles two tasks: 1) a segmentation sub-network aiming to generate the prostate segmentation, and 2) a voxel-metric learning sub-network aiming to improve the quality of the learned feature space supervised by a metric loss.
Specifically, the voxel-metric learning sub-network samples tuples (including triplets and pairs) in voxel-level through the intermediate feature maps. Unlike conventional deep metric learning methods that generate triplets or pairs in image-level before the training phase, our proposed voxel-wise tuples are sampled in an online manner and operated in an end-to-end fashion via multi-task learning. To evaluate the proposed method, we implement extensive experiments on a real CT image dataset consisting 339 patients. The ablation studies show that our method can effectively learn more representative voxel-level features compared with the conventional learning methods with cross-entropy or Dice loss. And the comparisons show that the proposed method outperforms the state-of-the-art methods by a reasonable margin.
\end{abstract}

}

\maketitle

\IEEEdisplaynontitleabstractindextext

%
\IEEEpeerreviewmaketitle


%
%
%
%

\begin{figure}[htbp]
  \centering
  \includegraphics[width=\linewidth]{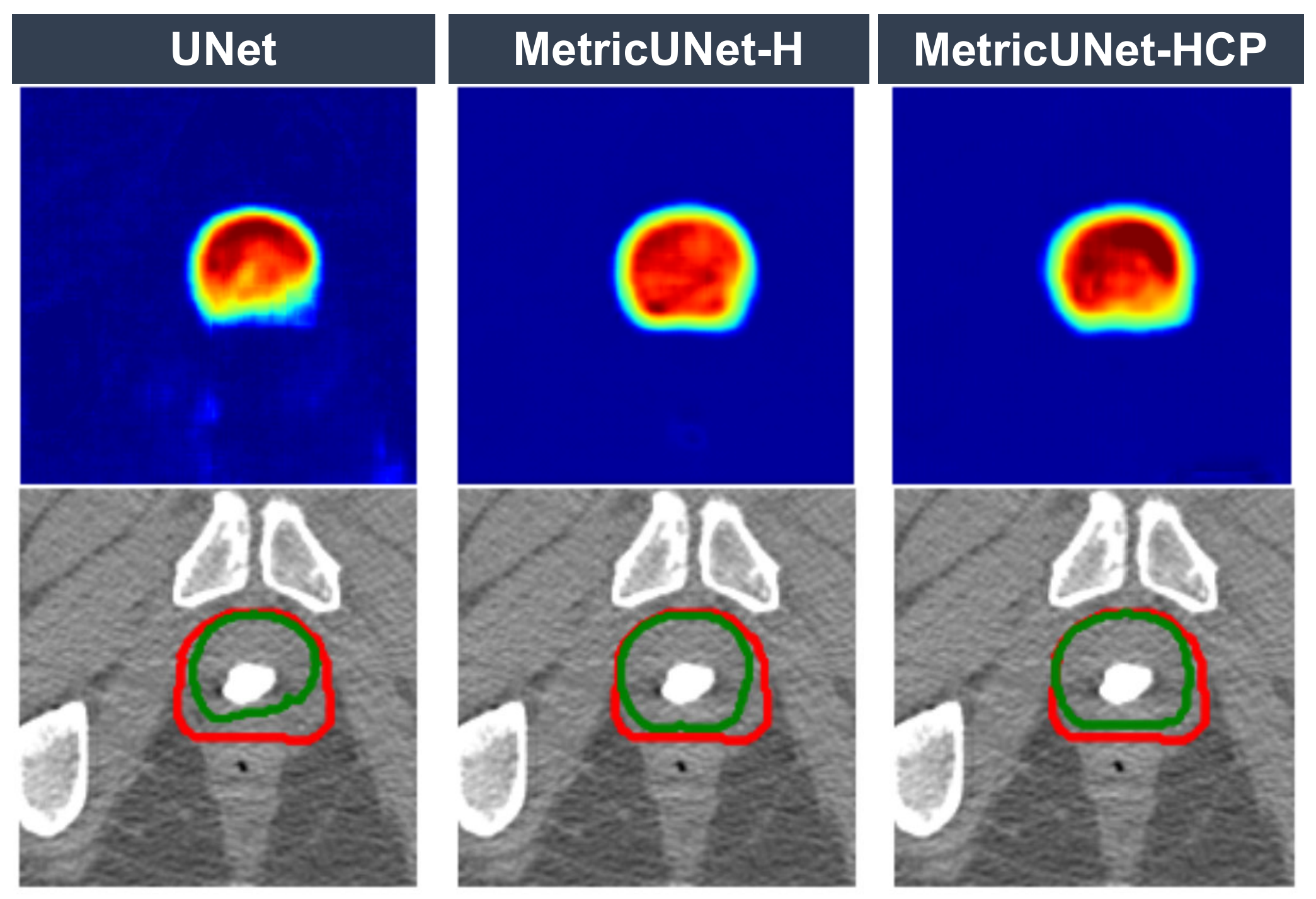}
  \caption{\label{fig:spotlight} Representative examples of heatmap and respective segmentation produced by UNets trained without and with our proposed online voxel-wise sampling (\ie, UNet versus MetricUNet-H and MetricUNet-HCP). Red contours denote the ground-truth, and green contours indicate the segmentation masks. UNet roughly identifies the prostate location in the heatmap. However, it failed to delineate the prostate boundary. In contrast, our MetricUNet-H with better learned voxel feature embeddings can generate precise segmentation in such challenging cases. Moreover, MetricUNet-HCP incorporates a contour-aware mechanism that can further delineate fuzzy prostate boundary more precisely.}
\end{figure}

\section{Introduction}

In the U.S., prostate cancer is the second most common cancer, and the second largest cause of cancer death for men. In 2020, about 33,330 patients died in about 191,930 newly developed prostate cancer cases. \cite{ps} The common treatment for prostate cancer is external beam radiation therapy (EBRT). In EBRT, precisely segmenting the prostate helps clinicians prepare dose planning. Automatic prostate segmentation by fully convolutional networks (FCNs) has achieved impressive results \cite{he2019pelvic,chen2020simple}. 
FCN and its variations, \eg, U-Net, V-Net, are state-of-the-art in pixel-to-pixel or voxel-to-voxel prediction tasks, \eg, segmentation \cite{long2015fully,johnson2016densecap,ronneberger2015u,milletari2016v,cciccek20163d,lian2018multi} and localization/detection \cite{dai2016r,bertinetto2016fully,lian2018hierarchical}, primarily because of the encoder-decoder architecture that gradually integrate global-to-local features.

\begin{figure}[thbp]
  \centering
  \includegraphics[width= \linewidth]{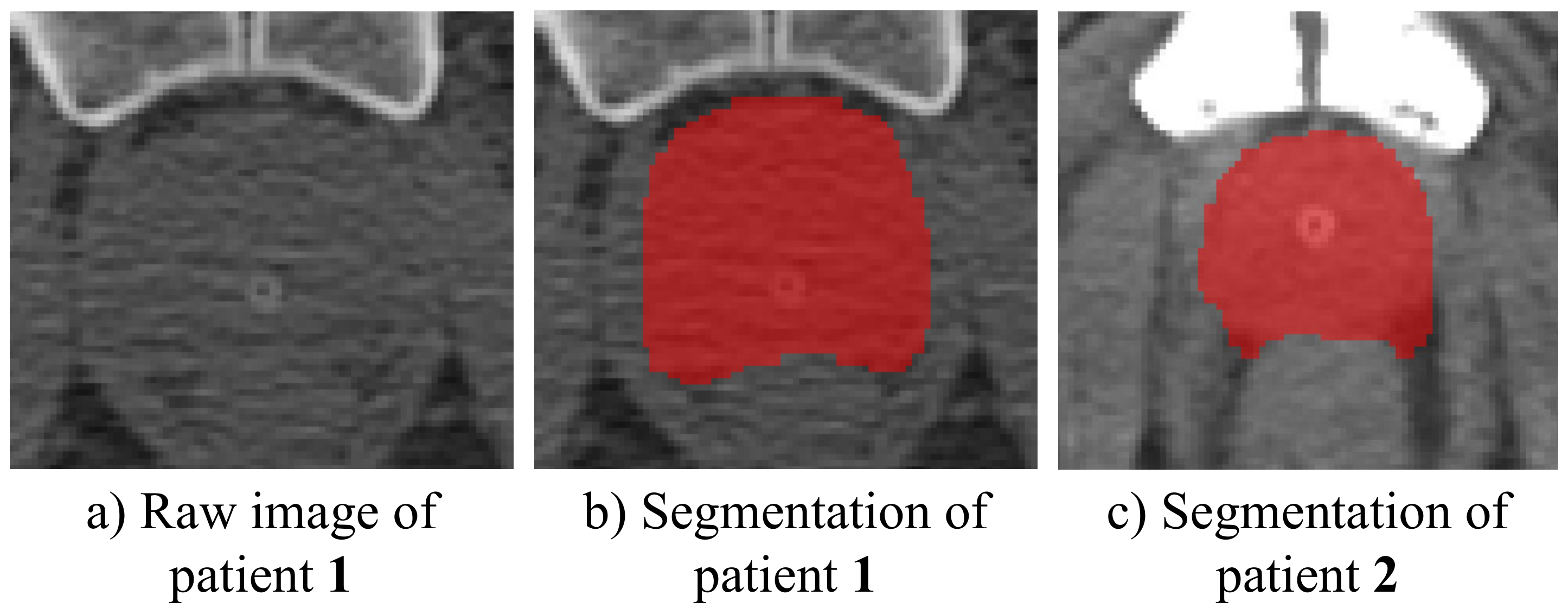}
  \caption{\label{fig:organvis} A raw CT slice and the ground-truth prostate segmentation produced by an experienced radiologist for two typical patients. The fuzzy organ boundary is showed by a) and b), and the large inter-subject variation is showed by b) and c).}
\end{figure}

However, existing methods typically train FCN by minimizing a loss which averages errors of predictions and ground-truth labels over all independent voxel locations, \ie, cross-entropy \cite{long2015fully,ronneberger2015u} or Dice \cite{milletari2016v} loss, without considering inter-voxel semantic correlations (regarding their labels) in the learned feature space.  
As a consequence, conventional FCN usually suffer from two common limitations: 
1) they can only generate rough probability map that indicates object location, but cannot produce finer segmentation for precisely delineating object boundaries \cite{DBLP:journals/pami/ChenPKMY18}; 2) they may generate incomplete segmentation with small fragmented pieces. 
These limitations are even more serious for prostate segmentation in computed tomography (CT) images, as shown by a representative example in Fig.~\ref{fig:spotlight}, since 1) CT images are often with low soft-tissue contrast, and 2) prostates usually have very unclear boundaries and also large variations across patients (See Fig. \ref{fig:organvis}).

These limitations indicate that the conventional FCNs are not adequate for pixel/voxel-level representation learning, leading to not refined segmentation results. And more powerful methods for learning pixel/voxel-level representations are of great interest.
To overcome these limitations, graph-based post-processing has been widely used to refine segmentation by FCNs. 
For example, \cite{DBLP:journals/pami/ChenPKMY18,trullo2017segmentation} adopted the fully connected conditional random fields (CRF) after FCN outputs to model inter-pixel/voxel relationships that generate more smooth segmentation by minimizing an energy function. 
The successes of these graph-based post-processing methods reveal the importance of explicitly modeling inter-voxel relationships in segmentation.
However, since post-processing is generally performed offline and does not contribute to voxel-wise feature learning, two independent stages may lead to sub-optimal segmentation results. Instead of solely using the cross-entropy/Dice loss followed by offline post-processing, or heuristically modifying the network architecture, we explore another direction, \ie, voxel-wise metric learning, to improve the feature quality.

\begin{figure*}[!t]
  \centering
  \includegraphics[width= \linewidth]{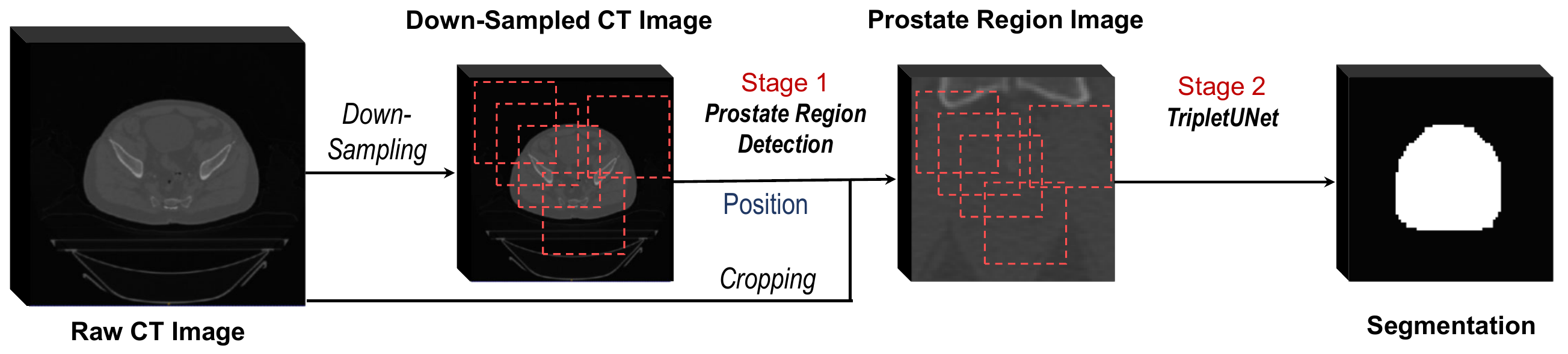}
  \caption{\label{fig:pipeline} The main stream of our proposed method. For simplicity, some complementary image processing procedures (\eg, body cropping, intensity and spatial normalization, training patch generation, etc.) are removed. The red dot boxes denote the random cropping strategy of the patches in training stage.}
\end{figure*}

We assume that, \emph{solely using the voxel-independent loss (\ie, cross-entropy or Dice loss) is not enough}, and \emph{synergistically modeling inter-voxel relationships along network training can learn much more discriminate feature space} for end-to-end image segmentation. 
Based on this assumption, a multi-task UNet (called \emph{MetricUNet}) is proposed in this paper and applied to segment prostates from raw (high-dimensional) CT images.
Fig.~\ref{fig:pipeline} shows the schematic diagram of our method, with two major stages.
Specifically, in Stage 1, we roughly locate prostate using a lightweight UNet trained with the down-sampled CT image. Then, in Stage 2, we crop each detected prostate region from the original CT image to train our MetricUNets to obtain precise prostate delineation. 

To improve voxel-level discriminative capacity of intermediate feature maps, we train MetricUNet by integrating \emph{online voxel-wise metric learning modules} for concurrent metric learning in voxel-embedding space. Notably, metric learning had been previously used in supervised image classification, retrieval tasks \cite{cheng2016person,parkhi2015deep}, and unsupervised feature learning \cite{chen2020simple}, and showed promising effectiveness in learning representative image-level feature embeddings. We model the metric learning scheme as an integration of triplet and contrastive learning in our method, which are two effective learning schemes for metric learning. Typically, triplet learning assumes semantically similar sample pairs are pulled closer together while dissimilar ones are pushed further apart. And contrastive learning assumes the similar sample pairs are close in the semantic embedding space.

In this paper, for segmentation, we regard \emph{voxels in each CT image as individuals}. However, it is challenging to incorporate the metric learning into voxel-level for segmentation. Since massive computation of voxel-level features through each iteration is unacceptable, we solved this problem by designing three complementary sampling strategies (\ie, random, focal hard negative and contour-aware sampling) to generate voxel-triplet and pairs for the learning of segmentation-oriented voxel-level feature embedding. We regard this sampling process is an essential step for incorporating metric learning into segmentation in voxel-level, as the computation resources are limited.
Our proposed method has shown three heuristics for improving segmentation: 
\begin{itemize}
\item we comprehensively model inter-voxel relationships by metric learning, thus driving the intermediate feature maps of the network to form a discriminate embedding space for more precise voxel-level classification (\ie, segmentation), with an example shown in Fig. \ref{fig:spotlight};
\item Our metric learning module is a general and cheap solution for precise segmentation, and can be easily inserted into any existing architectures without the need of any additional learnable weights or dramatically modify the network architecture;
\item The online voxel-metric learning module can be seamlessly combined with other strategies. For example, in the task of CT prostate segmentation, the module integrates contour-aware mechanism into sampling, so as to more precisely delineate the prostate boundary.
\end{itemize}

We have evaluated our proposed method on a real patient CT image dataset consisting of 339 subjects, leading to superior performance compared with the state-of-the-art methods for image segmentation. We also applied the proposed method on a benchmark dataset, \ie, PROMISE12, to evaluate its generalization ability. The competitive results further showed the robustness of the proposed method.

The rest of the paper is organized as follow. In Section 2, we introduce the related works for CT prostate segmentation and deep neural networks integrating metric-learning. In Section 3, we introduce the proposed methods in detail. In Section 4, we extensively evaluate our proposed method. Finally, in Section 5, we conclude the proposed method and discuss the future work.

\section{Related Work}
In this section, we review related work from two aspects: 1) existing automated methods developed for CT prostate segmentation, and 2) deep neural networks integrating metric learning.

\subsection{CT Prostate Segmentation}
Most recent studies for CT prostate segmentation fall into two categories, \ie, deformable model-based or learning-based methods.

Deformable model-based methods typically initialize a mesh, which is then optimized to move towards the object surface.
For example, \cite{shao2015locally} proposed a deformable model-based method which is jointly learned with a boundary regressor to include boundary information for guiding the prostate segmentation.
Considering that mesh initialization is a common challenge for deformable-based methods, \cite{gao2016accurate} proposed a context-based displacement regressor, which effectively eliminated bad initialization to improve the segmentation of male pelvic organs in CT images.
However, these methods are affected by the initialization and the contrast of the organ boundary, that limited its performance.

In learning-based methods, a classifier is typically trained to identify the label of a voxel (\eg, background or prostate) with hand-crafted features \cite{wu2006learning,mohamed2006deformable} or task-oriented features extracted by deep neural networks \cite{lian2018multi}. 
For example, \cite{wang2019hierarchical} proposed a FCN-based segmentation network incorporating the contour-aware strategy. Specifically, they leverage hierarchical virtual contour representations for identifying the fuzzy prostate boundaries in CT images. 
By taking the advantage of task-oriented learning and integration of local-to-global deep features, the FCN-based network largely improves the prostate segmentation performance in CT images. However, the automated segmentation produced by these methods are often incomplete with imprecise prostate boundaries. Therefore, several strategies have been explored to solve this problem. One is to incorporate additional guidance. For example, \cite{dong2019synthetic} proposed a framework that fuses multi-modality information for pelvic organ segmentation. Another way is using post-processing techniques for segmentation refinement. In another context of MRI prostate segmentation, \cite{wang2019deeply} added a contour refinement procedure after getting the prediction of prostate to smooth the segmentations.

\subsection{Metric Learning-based Deep Networks}
Metric learning is widely utilized with deep neural networks in classification and retrieval tasks, \eg, face recognition \cite{schroff2015facenet,parkhi2015deep,liu2017sphereface} and person Re-Identification (Re-ID) \cite{cheng2016person,ding2015deep}. 
Most applicable problems suffer from large intra-class sample variations. The situation is similar to the case of medical image segmentation.
For example, \cite{schroff2015facenet} proposed a CNN-based method that integrates online sampling of triplet image patches to learn representative feature embedding for face recognition and clustering.
\cite{wang2016joint} proposed to jointly learn from cross-image and single-image representations for person re-identification. Specifically, the network has three input branches to implement triplet-based learning.

However, all these existing methods perform triplet feature embedding at the image- or patch-level, \ie, they were not designed specifically for voxel-to-voxel prediction (\eg, segmentation). Notably, although a three-branch CNN was proposed in \cite{lim2018foreground} to perform segmentation using multi-scale images, this method is also patch-based without explicit consideration of inter-pixel/voxel relationships. In the context of medical image analysis, \cite{iesmantas2020enhancing} proposed a deep learning method with image-level metric embeddings for segmentation of cell nuclei in microscopy images.
This motivated us to train a FCN with online voxel-metric learning module to tackle the challenging task of CT prostate segmentation.

\section{Methods}

Prostate is located in a relatively small area in the whole pelvic CT image. Precisely segmenting prostate directly from the raw CT image is practically challenging and computationally infeasible, considering that the noisy background can strongly mislead the network.
Therefore, we design a two-stage deep learning framework to segmentation prostate in a coarse-to-fine fashion from raw CT images, as shown in Fig. \ref{fig:overall}. Specifically, the first stage quickly detects the prostate while the second stage performs fine prostate delineation.

\subsection{Prostate Detection}
In the first stage, a detection network is used to locate the region of the prostate. Unlike the conventional methods (\eg, the mass center \cite{ma2017combined}), which typically adopt statistical reference to roughly localize the prostate area, we follow our previous work \cite{he2019pelvic} to use a lightweight UNet architecture. Our lightweight detection network has a conventional UNet architecture but with a reduced filter number of $32$, to quickly and roughly segment the prostate on the down-sampled CT image. 

Specifically, we down-sample the images into 1/4 size of the original scale by trilinear interpolation.
Then, the light-weight UNet is trained with the 2D patches (size of $64\times 64 \times 5$ in this work) randomly cropped from the down-sampled CT image in a sliding-window fashion. 
After training, we feed the whole down-sampled CT images into the trained UNet to obtain the coarse prostate segmentation. This strategy avoids voting for patch-wise predictions.
The prostate region is localized with this coarse prostate segmentation by calculating the prostate centroid. The prostate centroid is calculated by an average of the reference prostate center and rough predicted prostate center. The reference prostate center is determined by the center of two femur heads. Hence, the rough predicted prostate center is determined by the coarse prostate segmentation.

Finally, we crop a fixed-sized prostate region centered on the calculated centroid from the raw CT image, and use it as the input for fine prostate segmentation in the subsequent stage. 
In this work, the size of the prostate region is set to $128\times128\times128$, which can cover the whole prostates for all training subjects.

\subsection{Fine Prostate Segmentation}

After getting the prostate region, in the second stage, we propose a multi-task UNet (\ie, denoted by MetricUNet-HCR) to generate fine prostate segmentation by introducing the online voxel-metric learning mechanism.

\begin{figure*}[!t]
  \centering
  \includegraphics[width=  \linewidth]{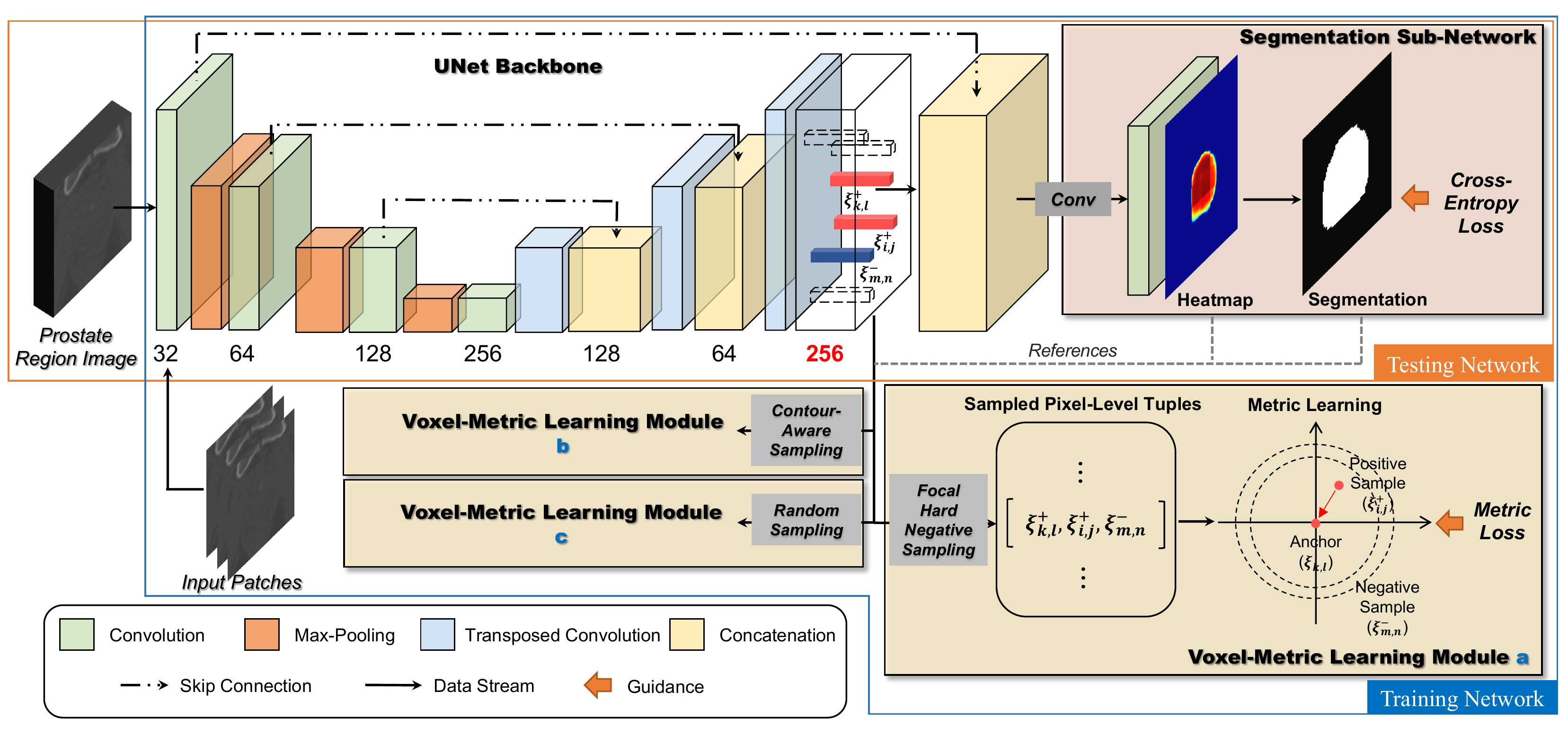}
    \caption{\label{fig:overall} The overall structure of the proposed MetricUNet-HCR (in Stage 2). The colored cubes denote the blocks with a specific operation layer (\ie, convolution, transposed convolution, pooling), and several complementary layers (\eg, convolutional layer, batch normalization layer, etc.). The details of the layers in different blocks are listed in Table \ref{Table:NetArc}. Please note that the segmentation sub-network and voxel-metric learning modules in different network configurations are of similar structures.}
\end{figure*}

\subsubsection{Network Architecture}

The architecture of an instantiation of our proposed MetricUNet, namely MetricUNet-HCR, is shown in Fig. \ref{fig:overall}. 
Similar to UNet \cite{ronneberger2015u}, our MetricUNet-HCR is also an encoder-decoder architecture with skip-connections for the extraction of voxel-level features that integrate local-to-global information. 
It consists of three down-sampling blocks, three up-sampling blocks, and an online voxel-metric learning module with three complementary sampling strategies. 
Specifically, MetricUNet-HCR has two sub-networks: 1) the segmentation sub-network for producing voxel-level predictions; 2) the voxel-metric learning modules for modeling the inter-voxel relationships via voxel-level feature embeddings.
The detailed architecture of the MetricUNet-HCR is shown in Table \ref{Table:NetArc} comprising a total number of parameters of 3.28 million.

\begin{table*}[htbp]
\renewcommand{\arraystretch}{1.3}
\centering
\caption{\label{Table:NetArc} Network architecture of the proposed MetricUNet-HCR. The 'Type' column lists the type of the layers, including, 'conv': convolutional layer; 'dconv': down-sampling with padding and matching convolutional layer; 'tconv': transposed convolutional layer; 'bn': batch normalization layer; 'r': ReLU layer; 'samp-h': focal hard negative sampling; 'samp-c': contour-aware sampling. 'Params' is formatted in \{kernel size, stride, padding\} for convolutional layers, \{kernel size, stride\} for pooling layers, and \{kernel size, stride\} for transposed convolutional layers. '\#' denote number of parameters.}
\begin{tabular}{cccccc}
\toprule[1pt]
\textbf{Block} & \textbf{Layer Name} & \textbf{Type} & \textbf{Input} & \textbf{Params} & \textbf{\#}\\
\toprule[1pt]
\multirow{3}{*}{conv block1} & conv1a & conv+bn+r & image & $3\times3\times32,1,1$ & $0.9K$\\
&conv1b & conv+bn+r & conv1a & $3\times3\times32,1,1$ & $9K$\\
& pool1 & max-pooling & conv1b & $2,2$ &-\\
\hline
\multirow{3}{*}{conv block2} & conv2a & conv+bn+r & pool1 & $3\times3\times64,1,1$ &$18K$\\
&conv2b & conv+bn+r & conv2a & $3\times3\times64,1,1$ & $37K$\\
& pool2 & max-pooling & conv2b & $2,2$ &-\\
\hline
\multirow{3}{*}{conv block3} & conv3a & conv+bn+r & pool2 & $3\times3\times128,1,1$ &$74K$ \\
&conv3b & conv+bn+r & conv3a & $3\times3\times128,1,1$ &$147K$\\
& pool3 & max-pooling & conv3b & $2,2$ &-\\
\hline
\multirow{3}{*}{conv block4} & conv3a & conv+bn+r & pool2 & $3\times3\times256,1,1$ &$295K$\\
&conv3b & conv+bn+r & conv3a & $3\times3\times256,1,1$ &$590K$\\
\hline
\multirow{4}{*}{upblock1} & upconv4a & tconv & dconv3 & $2, 2$&$131K$\\
& concate4b & concat & upconv4a, conv3b & - &-\\
&conv4c & conv+bn+r & concate4b & $3\times3\times128,1,1$ &$295K$\\
&conv4d & conv+bn+r & conv4c & $3\times3\times128,1,1$ & $147K$\\
\hline
\multirow{4}{*}{upblock2}& upconv5a & tconv & conv4c &  $2, 2$&$32K$\\
& concat5b & concat & upconv5a, conv2b  & - &-\\
& conv5c & conv+bn+r & concat5b & $3\times3\times64,1,1$ &$73K$\\
& conv5d & conv+bn+r & conv5c & $3\times3\times64,1,1$ &$37K$\\
\hline
\multirow{4}{*}{upblock3}& upconv6a & tconv & conv5d &  $2, 2$&$66K$\\
& concat6b & concat & upconv6a, conv1b & - &-\\
& conv6c & conv+bn+r & concat5b & $3\times3\times256,1,1$ &$664K$\\
& conv6d & conv+bn+r & conv6c & $3\times3\times256,1,1$ &$590K$\\
\hline
\multirow{3}{*}{Segmentation Sub-Network}& sega & conv+bn+r & conv6d &  $3\times3\times256,1,1$&$66K$\\
& segb & conv & sega & $1\times1\times2,1,0$ &$0.5K$\\
& lossce & loss & segb & Cross Entropy &-\\
\hline
\multirow{2}{*}{Voxel-Metric Learning Sub-Network a}& samph & samp-h & conv6d &  hard negative &-\\
& lossh & loss & samph & Metric Loss& -\\
\hline
\multirow{2}{*}{Voxel-Metric Learning Sub-Network b}& sampc & samp-c & conv6d &  contour-aware& -\\
& lossc & loss & sampc & Metric Loss& -\\
\hline
\multirow{2}{*}{Voxel-Metric Learning Sub-Network c}& sampr & samp-r & conv6d &  random &-\\
& lossr & loss & sampr & Metric Loss &-\\
\toprule[1pt]
\end{tabular}\\
\end{table*}

\subsubsection{Voxel-Metric Learning Module}

As discussed earlier, conventional FCNs typically ignore the inter-voxel relationship in the learned deep feature space. Thus, they may lead to fragmented segmentation. 
To tackle this challenge, we propose a voxel-metric learning module, by sampling heatmaps to voxel-wise tuples online. 

In this case, each voxel (along with the respective voxel-level feature representation) is a candidate to generate the tuples. 
The operation is performed on the feature maps from the last convolutional layer of the last up-sampling block. Formally, given a mini-batch of $N$ inputs, we denote the size of such feature maps as $N \times h \times w \times d$. Here, $d$ denote the feature length of the voxel. $h,w$ denote the feature height and width, respectively. As the voxel-level features may vary through each learning iteration, generating the whole set of tuples which has the quantities of $(N\times h \times w)^{(2\times3)}$ in each iteration is unacceptable. To effectively train the network, we select a subset of tuples in each iteration. Specifically, the selected tuples are first determined by $k$ anchors, for which we randomly select $m$ positive and negative samples. 

\begin{figure}[htbp]
  \centering
  \includegraphics[width=\linewidth]{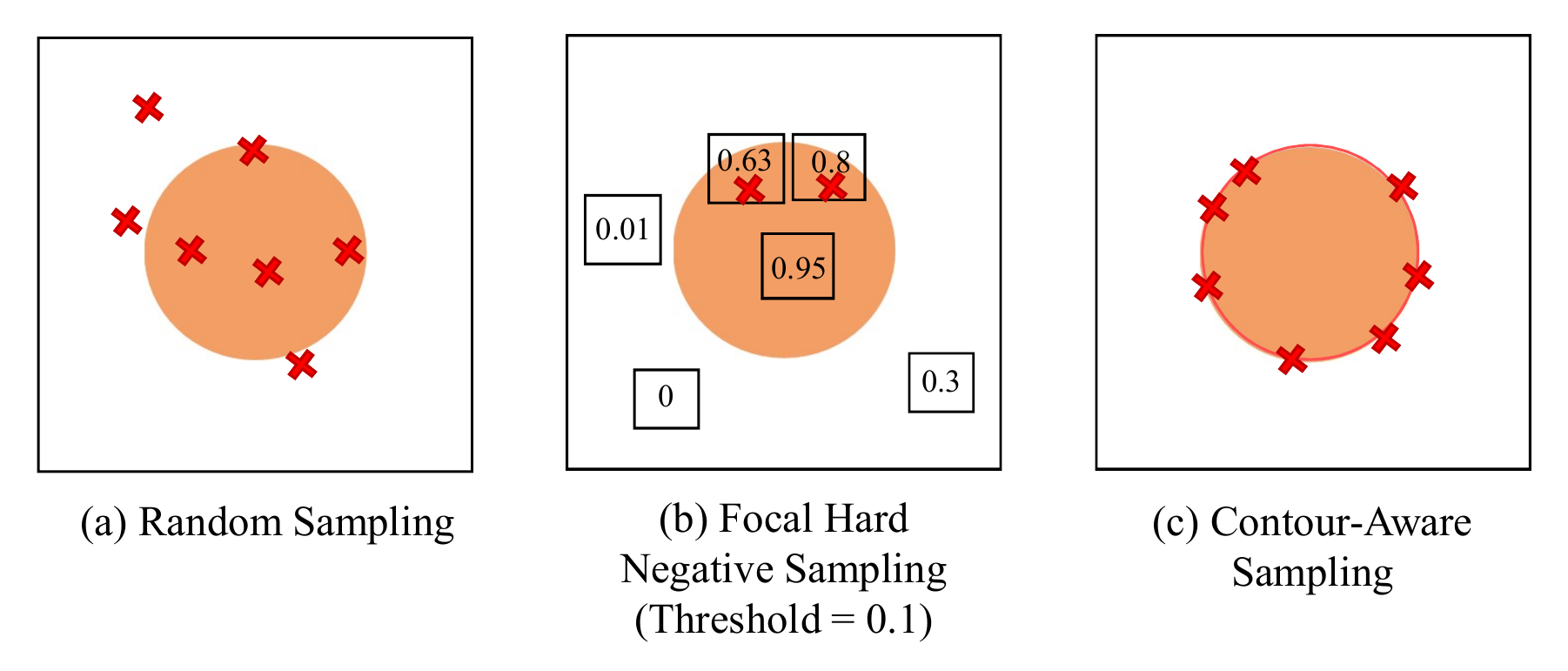}
    \caption{\label{fig:samp} A brief illustration of three different sampling strategies. Red cross denote the anchor point selected in one iteration. Please note that the box with value in (b) denote the predicted probability of the point in a certain position. The orange round and circle denoted the reference ground-truth of segmentation of the prostate which indicates the positive labels in the sampling scheme. The red circle in (c) denote the prostate contour and used for selecting anchor candidates.}
\end{figure}

As has been verified in previous work \cite{schroff2015facenet,duan2018deep}, different sampling strategies may strongly affect the performance of the network. 
Therefore, in this work, we explore three sampling strategies, \ie, 1) random sampling, 2) focal hard negative sampling, and 3) contour-aware sampling, and combine them for our MetricUNet. A brief illustration of these three sampling strategies is shown in Fig. \ref{fig:samp}. 
In the experiments, the latter two sampling strategies are proven to help network learn more discriminate features compare with random sampling.

\paragraph*{\textbf{i) Random Sampling}}

A basic and intuitive sampling method to produce the tuples is the random sampling. In random sampling method, the anchors are randomly selected in the voxels which held the positive labels (see in Fig. \ref{fig:samp} (a)). The positive and negative samples are also randomly selected according to their labels.

\paragraph*{\textbf{ii) Focal Hard Negative sampling}}

As proved by works in \cite{schroff2015facenet,cheng2016person}, the random sampling method may produce insufficient learned method, as the common tuples which satisfy the triplet and pair-wise distance assumption will dominate the training process.
Therefore, these methods mine only the hard/semi-hard samples.
The hard negative mining strategy used in conventional methods is typically implemented by selecting the triplets, which have an $\{$anchor,positive$\}$ pair with the maximum intra-class distance, and a $\{$anchor,negative$\}$ pair with the minimum inter-class distance. 
Obviously, the calculation of distances among all triplets or even a mini-batch is computationally expensive. 
Therefore, the work in \cite{schroff2015facenet} proposed to use triplets that violate the triplet distance assumption.

Inspired by the focal loss proposed in \cite{lin2017focal}, \emph{we assume that the hard negative samples can be revealed by the incorrect predictions}. Under this assumption, we propose a focal hard negative sampling strategy, which adopts the heatmap generated by the segmentation sub-network as the reference to find the anchors. 
Specifically, for each voxel, we first calculate the uncertainty of the prediction, which is implemented as the absolute difference between the prediction and its ground-truth label. Then, a threshold $\tau$ is applied to select the bad predictions with large differences as the hard samples. The operation can be written as:
\begin{align}
\begin{split}
s^{hard}_{ij} = \left\{ \begin{aligned}
 &~1  &~~~~~~~~~~~~&\text{if} |(\hat{y}_{ij}-y_{ij})| > \tau \\
 &~0  &~~~~~~~~~~~~&\text{if} |(\hat{y}_{ij}-y_{ij})| \leq \tau
\end{aligned}
\right.
\label{eq:Hard}
\end{split}
\end{align}

where $s^{hard}_{ij}=1$ denote that the pixel at point $(i,j)$ is labeled as a hard negative sample. The threshold $\tau$ in this work is set to 0.1. This means both the semi-hard (\ie, near threshold) samples and hard (\ie, far away from threshold) samples are used to construct the tuples. 
Finally, as shown in Fig. \ref{fig:samp} (b), the intersection of hard sample map $s$ and positive sample map $p$, \ie, the positive hard sample map, is used as anchors to generate the tuples.

\paragraph*{\textbf{iii) Contour-Aware sampling}}
For organ (\eg, prostate and gland) segmentation in medical images, the organ boundaries are often clinically important, but hard to be distinguished. Moreover, FCNs are not capable to directly generate refined segmentation for blurry organ boundaries. Therefore, various works leverage the contour-aware strategy to enhance the boundary discriminate ability of FCN. Typically, the contour-aware strategy forces the network to give special focuses on the contours/boundaries of the target organ, by introducing either specific designed loss terms \cite{zhou2019cia} or guidance \cite{DBLP:journals/mia/ChenQYDQH17,he2019pelvic} to the network. Inspired by the successes of these existing methods, we propose a special sampling strategy to incorporate contour-awareness, \ie, the contour-aware sampling, into voxel-wise feature embedding. Specifically, in each training iteration, we randomly select a number of points located on the prostate boundary as anchors, and construct tuples via these contour-aware anchors (See in Fig. \ref{fig:samp}.(c)). The positive samples are also randomly selected according to the contour points. The negative samples are randomly selected from the voxels with negative labels.

\subsubsection{Multi-Task Learning}
 
 As the whole network is constructed of multiple branches with multiple guidance, we leverage the power of multi-task learning to train them simultaneously. 
 
The voxel-wise metric learning is performed after getting the hard-negative and contour-aware sampling.
Given one triplet as $\{\xi_{k,l}, \xi_{i,j}, \xi_{m,n}\}$, we denote $\xi_{k,l}$ as the feature representation of the $(k,l)$th point (\ie, an anchor) on the feature map $\xi$. Similarly, $\xi_{i,j}$ is the feature representation of the $(i,j)$th point which has the same label as the anchor, while $\xi_{m,n}$ has the opposite label. The loss function $\mathcal{L}_{metric}$ for voxel-wise feature embedding is then defined as

\begin{align}
\begin{split}
&\mathcal{L}_{metric} = \\
& \sum_{n=1}^{N}\sum_{k,l \in K}\sum_{i,j \in J}\sum_{m,n \in M}{[(\underbrace{||\xi_{k,l}- \xi_{i,j}||_2^2 
 - ||\xi_{k,l} - \xi_{m,n}||_2^2 + \sigma}_{\text{triplet loss}})} \\ 
& + \beta\sum_{n=1}^{N}\sum_{k,l \in K}\sum_{i,j \in J}{(\underbrace{||\xi_{k,l}- \xi_{i,j}||_2^2 + \epsilon}_{\text{positive pair-wise loss}})}]
\label{eq:Trip}
\end{split}
\end{align}

where $K$ denote the anchor set, $J$ denote the voxel set of positive labels, $M$ denote the voxel set of negative labels. 
Specifically, $\mathcal{L}_{metric}$ consists of two terms balanced by a hyper-parameter $\beta$. The first term is the triplet loss term to control the distances between positive samples and negative samples. The second term is an pair-wise loss term to control the cluster of the intra-class positive samples. 
The hyper-parameter $\sigma$ denote the margin to control the distance between positive pairs and negative pairs, and $\epsilon$ denote the maximum distance of the positive pairs. In this paper, we set the hyper-parameters $\sigma=0.7$, $\epsilon=0.01$ and $\beta=0.1$. 

Concurrently, the segmentation sub-network is trained by minimizing the cross-entropy loss:
\begin{align}
\begin{split}
&\mathcal{L}_{CE} = - \frac{1}{N} \sum \limits^{N}_{n = 1} log(\hat{p}_n, l_n)
\label{eq:CE}
\end{split}
\end{align}

Considering that distance-based losses (\eg, triplet loss in Eq.(\ref{eq:Trip})) are of large magnitude, we use a factor $\{\lambda=0.01\}$ to balance these two kind of losses (\ie, cross-entropy loss and metric loss). Therefore, the final loss can be formally written as,
\begin{equation}
\begin{array}{rrclcl}
\displaystyle \mathcal{L} = \mathcal{L}_{CE} + \lambda \sum_{t=1}^{T}\mathcal{L}_{metric}
\end{array}
\label{eq:Ratio}
\end{equation} 
where $t$ denote the $t$th sampling strategy. 

Notably, in the training phase, the network is optimized with two learning paths, \ie, the segmentation path and the metric learning path. In the testing phase, the voxel-wise prediction is directly output by the segmentation path, without the voxel-metric learning module.

\section{Experimental Results}

\subsection{Data Setup}

We evaluate our method on a large planning CT image dataset containing 339 patient CT scans. The images were collected by North Carolina Cancer Hospital, and the prostate contours manually delineated by two clinicians are adopted as the ground-truth labels. The image size is $512\times512\times(61\sim508)$, with in-plane resolution as $0.932\sim1.365$mm, and slice thickness as $1\sim3$mm. As the images were collected in a long period using different scanners, the image size and resolution are varied across the patients. Automatic segmentation on this dataset is challenging because: 1) the images include different patient positions and are of different sizes; 2) one patient only has one image in this planning image dataset. We randomly partition the dataset into $70\%$ for training, $10\%$ for validation and $20\%$ for testing.

\begin{table*}[htbp]
\renewcommand{\arraystretch}{1.3}
\centering
\caption{\label{Table:Config} Different network configurations of the proposed MetricUNet (including conventional UNet as baseline).}
\setlength{\tabcolsep}{8pt}
\begin{tabular}{cccccccc}
\toprule[1pt]
\multirow{2}{*}{\textbf{Method}} & \multicolumn{3}{c}{\textbf{Sampling Strategy}} &  & \multicolumn{3}{c}{\textbf{Loss}}  \\
\cline{2-4}
\cline{6-8}
&\emph{Random}&\emph{Hard Negative}&\emph{Contour-Aware}&& \emph{Cross-Entropy} & \emph{Positive Pair} & \emph{Triplet} \\
\toprule[1pt]
\textbf{UNet (Baseline)} & - & - & - &  & \checkmark & - & - \\
\textbf{UNet++} & - & - & - &  & \checkmark & - & - \\
\textbf{MetricUNet-R-Sep} & \checkmark & - & - &  & \checkmark & - & \checkmark \\
\hline
\textbf{MetricUNet-R} & \checkmark & - & - &  & \checkmark & - & \checkmark \\
\textbf{MetricUNet-H} & - & \checkmark & - &  & \checkmark & - & \checkmark \\
\textbf{MetricUNet++-H} & - & \checkmark & - &  & \checkmark & - & \checkmark \\
\textbf{MetricUNet-C} & - & - & \checkmark &  & \checkmark & - & \checkmark \\
\hline
\textbf{MetricUNet-HR} & \checkmark & \checkmark & - &  & \checkmark & - & \checkmark \\
\textbf{MetricUNet-HC} & - & \checkmark & \checkmark &  & \checkmark & - & \checkmark \\

\hline
\textbf{MetricUNet-HP} & - & \checkmark & - &  & \checkmark & \checkmark & \checkmark \\
\textbf{MetricUNet-HRP} & \checkmark & \checkmark & - &  & \checkmark & \checkmark & \checkmark \\
\textbf{MetricUNet-HCP} & - & \checkmark & \checkmark &  & \checkmark & \checkmark & \checkmark \\

\toprule[1pt]
\end{tabular}
\end{table*}

\subsection{Implementation Details}
Our method was implemented using the popular open-source framework \emph{PyTorch} \cite{paszke2017automatic}. The training of the proposed network is accelerated by four NVidia GTX 1080Ti GPUs. 

We only did necessary image pre-processing on the original CT scans. First, we used trilinear interpolation to normalize the image spacing into $1\times1\times1mm^3$. Then, to eliminate the inf luence caused by singular values, we normalized the image intensities to $[0,255]$. 

The images with body regions were cropped by a simple threshold-based method to reduce the black background noise. Then, a sliding window method was implemented on these cropped body image parts to generate 2D training patches. As the 2D-based network ignores learning the inter-slice correlations, we used image patches with continuous slices as the inputs to enhance this ability. In this case, multiple slices were used to predict the label for the middle slice. In training, the MetricUNet takes size of $64\times64\times3$ patches as the inputs in the second stage. 
In the testing phase, the predictions of the region images are directly obtained from the network.
This strategy can better incorporate inter-slice context for 2-D based methods. To construct training data, we cropped $500$ patches from each training image. In testing, we directly apply the trained network to the cropped and down-sampled body region images (in the first stage) or the organ region images (in the second stage) to avoid voting for the patch-wise predictions. The goal of the experiments is to verify the effectiveness of our proposed voxel-metric learning method in improving segmentation performance and therefore we deployed our model using 2-D architectures (instead of 3-D) to obtain viable and comparable results for all architectures is shorter amount of time.

The networks were trained with the batch size of $30$ on each GPU. All the competitive networks were optimized by standard Stochastic Gradient Descent (SGD) algorithm. The learning rate was decayed from $0.01$ by the 'Poly' decay method.

\begin{figure*}[htbp]
  \centering
  \includegraphics[width= 0.48\linewidth]{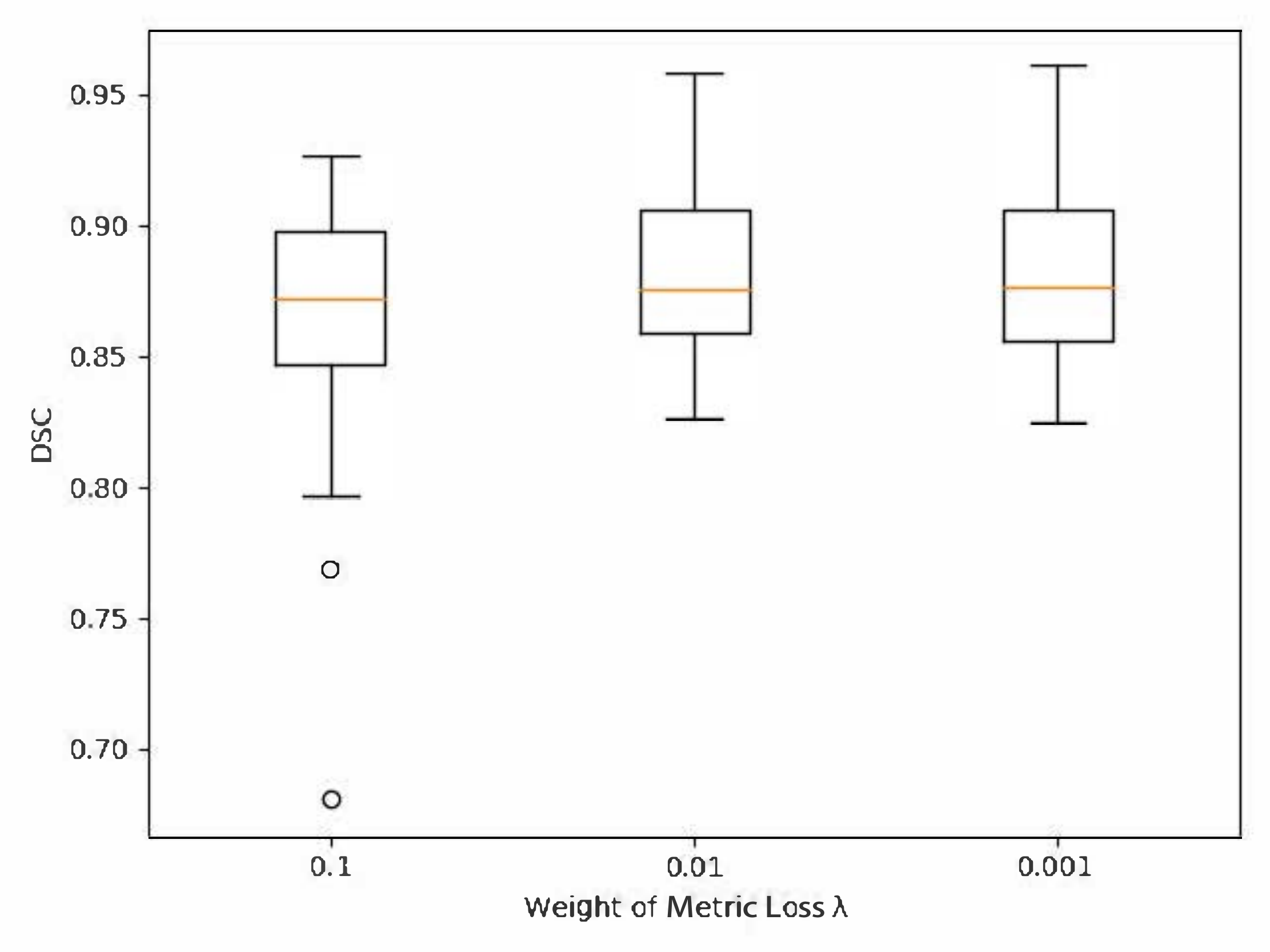}
  \includegraphics[width= 0.48\linewidth]{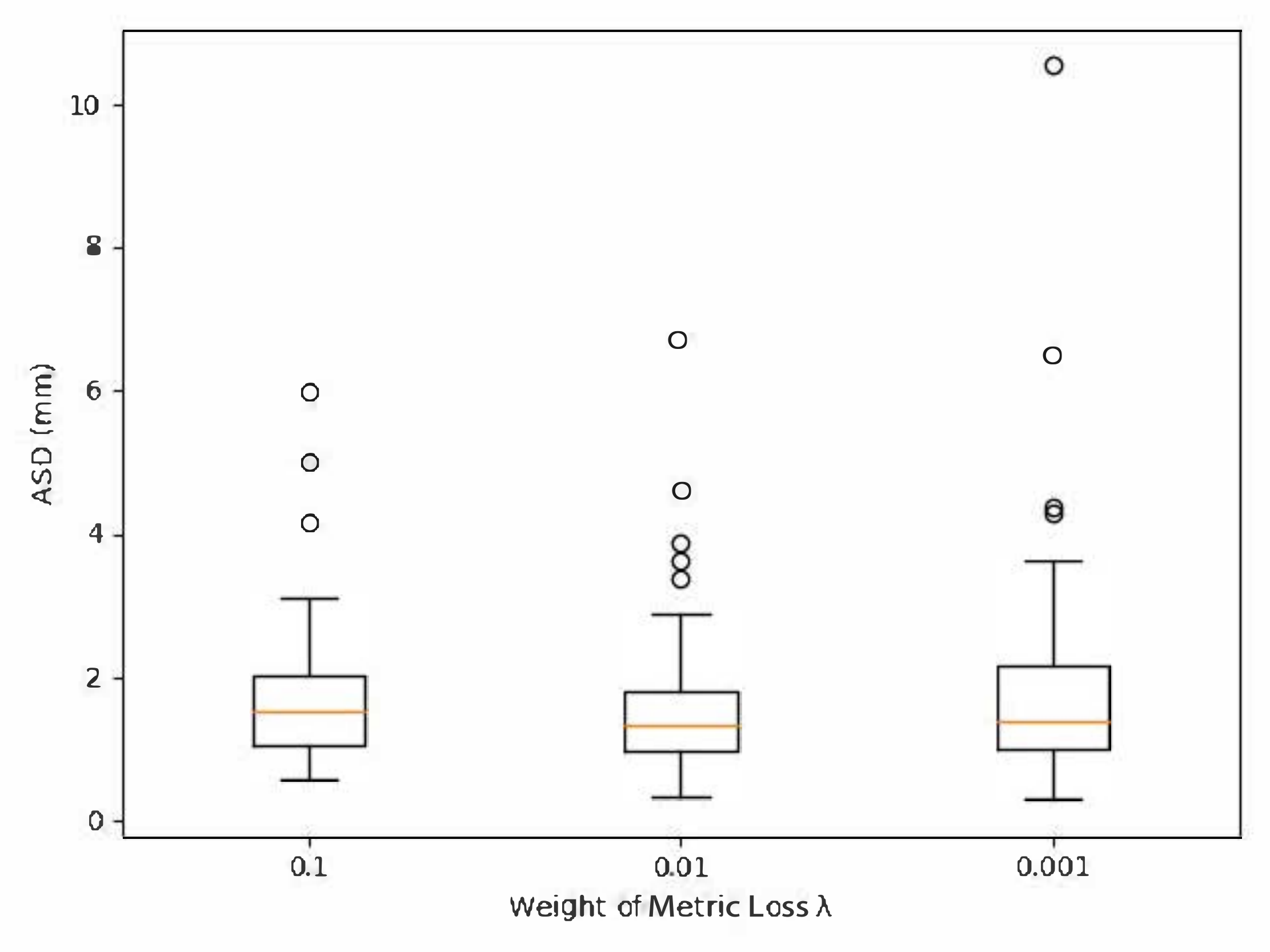}\\
  \includegraphics[width= 0.48\linewidth]{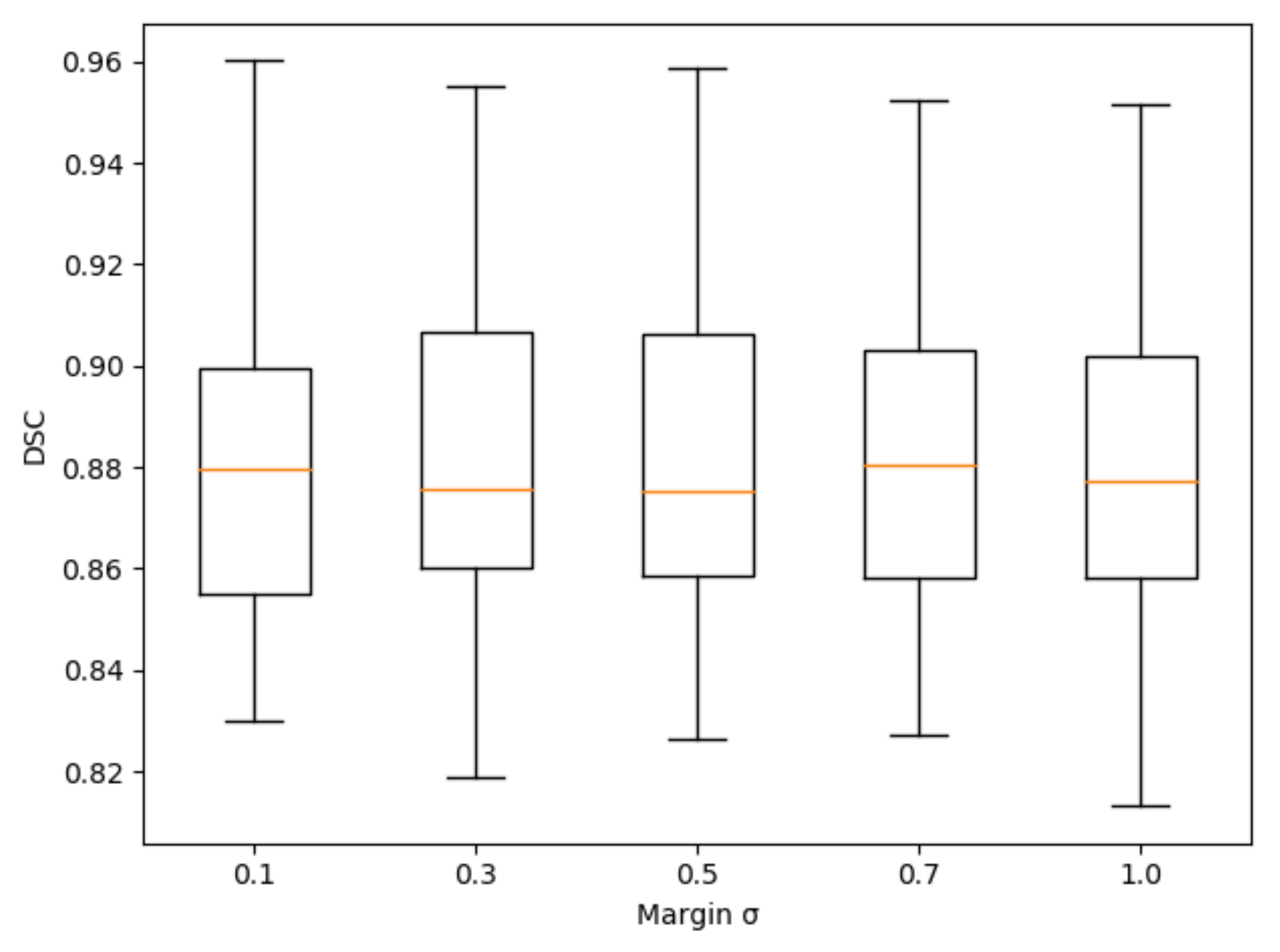}
  \includegraphics[width= 0.48\linewidth]{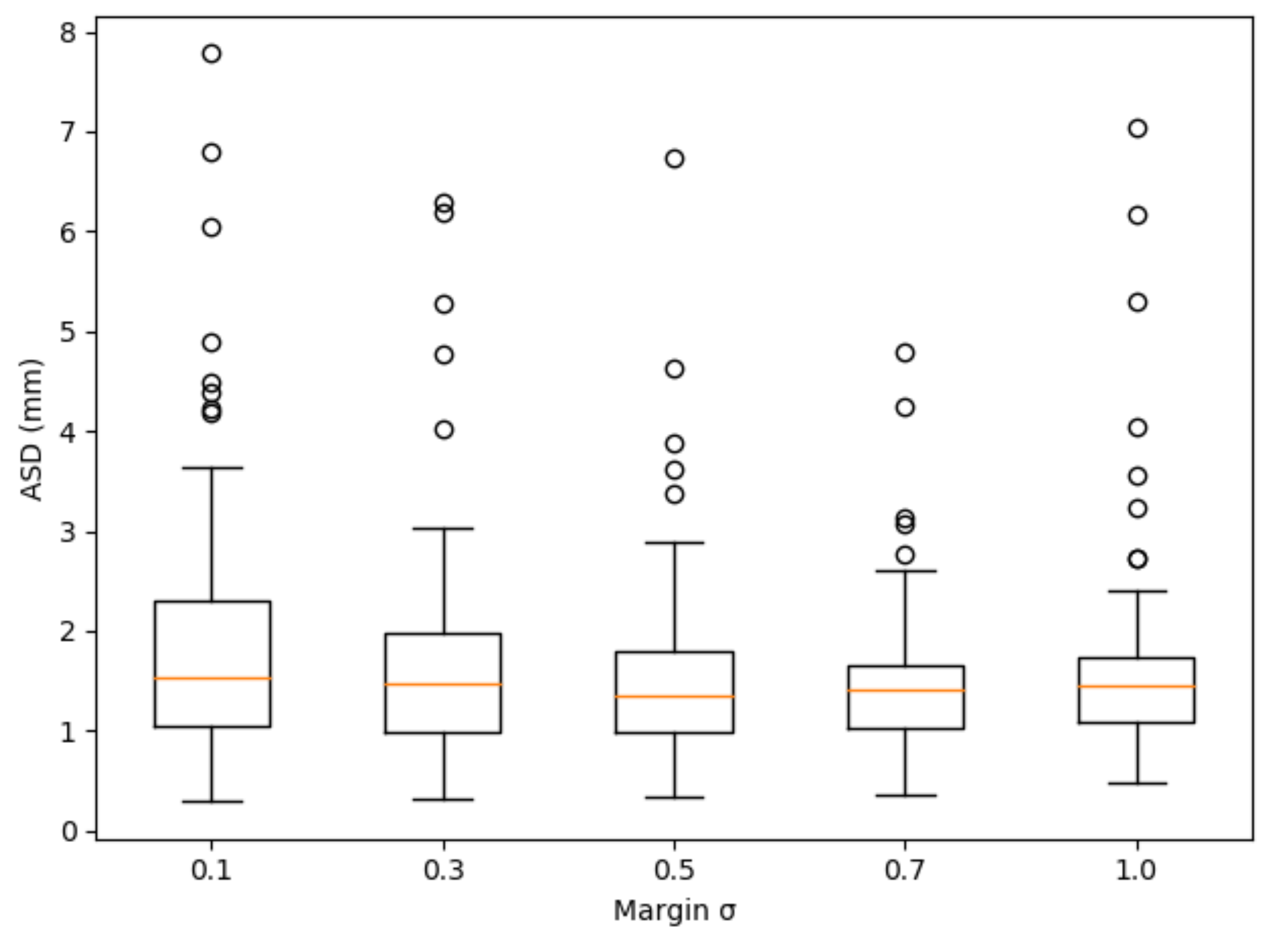}
  \caption{\label{fig:comp_alpha} The performance comparison in DSC and ASD of MetricUNet-R as a function of weight $\lambda$ and margin $\sigma$.}
\end{figure*}

\begin{table*}[htbp]
\renewcommand{\arraystretch}{1.3}
\centering
\caption{\label{Table:samplingstrategy} Quantitative comparison of segmentation performance with MetricUNets in DSC, ASD(mm), SEN and PPV of different network configurations. All the networks are trained with the following hyper-parameters: $k=20$, $m=1$, $\sigma=1.0$ and $\lambda=0.01$.}
\setlength{\tabcolsep}{2pt}
\begin{tabular}{cccccccccccc}
\toprule[1pt]
\multirow{2}{*}{\textbf{Method}} & \multicolumn{2}{c}{\textbf{DSC}} && \multicolumn{2}{c}{\textbf{ASD}} && \multicolumn{2}{c}{\textbf{SEN}} && \multicolumn{2}{c}{\textbf{PPV}} \\
\cline{2-3}
\cline{5-6}
\cline{8-9}
\cline{11-12}
&Mean $\pm$ std & Median && Mean $\pm$ std & Median && Mean $\pm$ std & Median && Mean $\pm$ std & Median \\
\toprule[1pt]
\textbf{UNet (Baseline)} & 0.8376 $\pm$ 0.0633 & 0.8501 && 3.8922 $\pm$ 2.9335& 2.9717 && 0.8733 $\pm$ 0.0777& 0.8819 & & 0.8143 $\pm$ 0.0918  & 0.8246\\
\textbf{MetricUNet-R-Sep} & 0.8514 $\pm$ 0.0506 & 0.8550 && 3.7606 $\pm$ 2.7031 & 3.2781 && 0.8497 $\pm$ 0.0829 & 0.8584 && 0.8616 $\pm$ 0.0613 & 0.8717 \\
\textbf{UNet++} & 0.8482 $\pm$ 0.0499 & 0.8549 && 1.9534 $\pm$ 1.0396 & 1.6059 && 0.8979 $\pm$ 0.0625 & 0.9209 && 0.8152 $\pm$ 0.0942 & 0.8240 \\
\hline
\textbf{MetricUNet-R} & 0.8817 $\pm$ 0.0304 & 0.8784 && 1.4619 $\pm$ 0.7715 & 1.2683 && 0.8727 $\pm$ 0.0588 & 0.8700 && 0.8952 $\pm$ 0.0449 & \textbf{0.9024} \\
\textbf{MetricUNet-H} & 0.8628 $\pm$ 0.0407 & 0.8951 && 1.7695 $\pm$ 1.1246 & 1.3915 && 0.8765 $\pm$ 0.0582 & 0.8811 && 0.8897 $\pm$ 0.0440 & 0.8940 \\
\textbf{MetricUNet++-H} & 0.8657 $\pm$ 0.0370 & 0.8689 && 1.7364 $\pm$ 0.8511 & 1.4176 && 0.9041 $\pm$ 0.0661 & 0.9271 && 0.8376 $\pm$ 0.0595 & 0.8428 \\
\textbf{MetricUNet-C} & 0.8822 $\pm$ 0.0285 & 0.8775 && 1.6355 $\pm$ 1.0693 & 1.2944 && 0.8732 $\pm$ 0.0561 & 0.8697 && \textbf{0.8955 $\pm$ 0.0437} & 0.8996 \\
\hline
\textbf{MetricUNet-HR} & 0.8822 $\pm$ 0.0335&0.8776&& 1.4295 $\pm$ 0.8719 & \textbf{1.2498} & & 0.8760 $\pm$ 0.0631&0.8777&& 0.8931 $\pm$ 0.0444 & 0.8974\\
\textbf{MetricUNet-HC} & 0.8804 $\pm$ 0.0315&0.8750&& 1.3987 $\pm$ 0.5836 & 1.2721 & & 0.8714 $\pm$ 0.0617&0.8768&& 0.8947 $\pm$ 0.0475 & 0.8932\\

\hline
\textbf{MetricUNet-HP} & 0.8811 $\pm$ 0.0303 & 0.8790 && 1.5274 $\pm$ 0.7687 & 1.3110 && 0.8719 $\pm$ 0.0602 & 0.8694 && 0.8955 $\pm$ 0.0474 & 0.8992 \\
\textbf{MetricUNet-HRP} & 0.8834 $\pm$ 0.0302 & 0.8814 && 1.5701 $\pm$ 0.9437 & 1.3329 && 0.8804 $\pm$ 0.0592 & \textbf{0.8842} && 0.8907 $\pm$ 0.0433 & 0.8953 \\
\textbf{MetricUNet-HCP} & \textbf{0.8839 $\pm$ 0.0317} & \textbf{0.8828} && \textbf{1.3907 $\pm$ 0.6651} & 1.2575 && \textbf{0.8805 $\pm$ 0.0571}& 0.8798 && 0.8914 $\pm$ 0.0452 & 0.8923 \\

\toprule[1pt]
\end{tabular}\\
\end{table*}

\subsection{Metrics}

We use five commonly used metrics to evaluate the performance of the proposed method: Dice Similarity Coefficient (DSC), Average Surface Distance (ASD), Positive Predictive Value (PPV), Sensitivity (SEN), and 95th percentile of Hausdorff distance (HD95). Let $Vol_{seg}$ denote the volume of a prediction, $Vol_{gt}$ denote the volume of the ground-truth segmentation, $Sur_{seg}$ denote the surface of a prediction, $Sur_{gt}$ denote the surface of ground-truth, and $d(a,b)$ denote the Euclidean distance between points $a$ and $b$. The five metrics can be written as,

\noindent(1) Dice Similarity Coeffient (DSC):
\begin{equation}
\begin{array}{rrclcl}
\displaystyle DSC=\frac{2\|Vol_{gt}\cap Vol_{seg}\|}{\|Vol_{gt}\|+\|Vol_{seg}\|};
\end{array}
\label{eq:DSC} 
\end{equation}

\noindent(2) Average Surface Distance (ASD):
\begin{equation}
\begin{array}{rrclcl}
\displaystyle ASD=\frac{1}{2}\{\mathop{mean}\limits_{a\in Sur_{gt}}\mathop{min}\limits_{b\in Sur_{seg}}{d(a,b)} \\
        + \mathop{mean}\limits_{a\in Sur_{seg}}\mathop{min}\limits_{b\in Sur_{gt}}{d(a,b)} \};
\end{array}
\label{eq:ASD} 
\end{equation}

\noindent(3) Positive Predictive Value (PPV) and Sensitivity (SEN):
\begin{equation}
\begin{array}{rrclcl}
\displaystyle PPV=\frac{\|Vol_{gt} \cap Vol_{seg}\|}{\|Vol_{seg}\|};
\end{array}
\hspace{1em}
\begin{array}{rrclcl}
\displaystyle SEN=\frac{\|Vol_{gt} \cap Vol_{seg}\|}{\|Vol_{gt}\|};
\end{array}
\label{eq:PPV} 
\end{equation}

\noindent(4) 95th percentile of Hausdorff distance (HD95):

The definition of Hausdorff distance (HD) is,
\begin{equation}
\begin{array}{rrclcl}
\displaystyle HD=\mathop{max}\{\mathop{max}\limits_{a\in Sur_{gt}}\mathop{min}\limits_{b\in Sur_{seg}}{d(a,b)}, \\
        \mathop{max}\limits_{a\in Sur_{seg}}\mathop{min}\limits_{b\in Sur_{gt}}{d(a,b)}\}.
\end{array}
\label{eq:HD95} 
\end{equation}

\begin{table*}[htbp]
\renewcommand{\arraystretch}{1.3}
\centering
\caption{\label{Table:sota} Quantitative comparison of segmentation performance with MetricUNets and state-of-the-art methods. * denote the results are reported on the 339 prostate dataset with the same data split. 'DM' denote deformable model, 'RF' denote random forest, 'DNN' denote deep neural network, 'GAN' denote generative adversarial network.}
\setlength{\tabcolsep}{2pt}
\begin{tabular}{cccccccc}
\toprule[1pt]
\textbf{Method} & \textbf{Type} & \textbf{Num. Cases}  & \textbf{DSC}  & \textbf{SEN} & \textbf{PPV} & \textbf{ASD} & \textbf{HD95} \\
\toprule[1pt]
\textbf{\cite{costa2007automatic}} & DM &16&-&0.75&0.80& - & -\\
\textbf{\cite{shao2015locally}} & DM + RF & 70 &0.88$\pm$0.02& - & - &1.86$\pm$0.21 & -\\
\textbf{\cite{gao2016accurate}} & DM + RF & 313 &0.87$\pm$0.04 &0.88$\pm$ - &0.85$\pm$ - &1.77$\pm$0.66 & -\\
\hline
\textbf{\cite{dong2019synthetic}-CT} & DNN & 140 
&0.82$\pm$0.09
&0.83$\pm$0.07
&0.87$\pm$0.05
&-
&8.52$\pm$1.09
\\
\textbf{\cite{dong2019synthetic}-CT+MRI} & DNN+GAN & 140 &0.87$\pm$0.04
&0.86$\pm$0.08
&0.88$\pm$0.08
&-
&6.35$\pm$3.11\\
\textbf{\cite{ronneberger2015u}*} & Cascaded DNN & 339 &0.84$\pm$0.06
&0.87$\pm$0.08
&0.81$\pm$0.09 
&3.89$\pm$2.93
&15.60$\pm$21.51\\
\textbf{\cite{milletari2016v}*} & Cascaded DNN & 339 
&0.85$\pm$0.04
&0.88$\pm$0.06
&0.84$\pm$0.07
&2.27$\pm$1.16
&12.36$\pm$15.26\\
\textbf{\cite{zhou2018unet++}*} & Cascaded DNN & 339 
&0.85$\pm$0.04
&0.89$\pm$0.06
&0.81$\pm$0.09
&1.95$\pm$1.16
&10.95$\pm$16.52\\
\textbf{\cite{he2019pelvic}*} & Cascaded DNN & 339 
&0.87$\pm$0.03
&0.88$\pm$0.05
&0.87$\pm$0.05
&1.71$\pm$1.01
&7.12$\pm$9.95\\
\textbf{\cite{nnUNet}(nnUNet)*} & Cascaded DNN & 339 
&0.87$\pm$0.06
&0.86$\pm$0.07
&0.88$\pm$0.08
&1.61$\pm$1.05
&6.64$\pm$6.54\\
\hline
\textbf{MetricUNet-HCP}* & Cascaded DNN & 339 &\textbf{0.88$\pm$0.03}
&\textbf{0.88$\pm$0.05}
&\textbf{0.89$\pm$0.04}
&\textbf{1.39$\pm$0.66}
&\textbf{5.24$\pm$3.21}\\
& \textbf{p-value}(vs. nnUNet) & - & 0.014 & 0.042 & 0.051 & 0.014 &  0.015 \\
\toprule[1pt]
\end{tabular}
\end{table*}

\subsection{Ablation Study}

To analyze the importance of different components of our MetricUNet, we did comprehensive experiments to evaluate: 1) the effectiveness of voxel metric learning, and 2) the influence of different sampling strategies, and 3) the effectiveness of the number of samples per iteration. Please note that, the aim of the ablation study is to verify the proposed metric learning method. Therefore, for the convenience of perform the experiments, we use UNet as the backbone. For the baseline method, we construct UNet to segment on the prostate region image which is same as the proposed two-stage learning framework. For a fair comparison, we avoid using many complicated techniques for training of the deep networks, \eg, data augmentation, dropout, residual block, dense block, etc., for which can improve the segmentation performance of the network, with the aim to illustrate the effectiveness of the proposed metric learning method. In this paper, we compose MetricUNets with six different configurations, with different combination of sampling strategies, architectures and losses. To evaluate the generalization ability of the proposed method, we also utilizing UNet++ as the backbone, thus constructing MetricUNet++ and MetricUNet++-H methods. The setting of these networks in detail is listed in Table \ref{Table:Config}. The second set of row introduces the proposed networks trained by tuples from one sampling methods. The third set of row introduces the proposed networks trained by tuples from two sampling methods jointly. The suffix 'Sep' means the network is trained separately by the two losses, where the layer before the sampling layer is only trained under the guidance of metric loss, and the layer after that is trained under the guidance of classification loss (\ie, the cross-entropy loss). We construct this network to evaluate the performance of pure-metric learning.

\subsubsection{The effectiveness of voxel-metric learning for segmentation}
Firstly, we verified the effectiveness of the proposed voxel-metric learning method from two aspects: 1) the converge rate, and 2) the segmentation accuracy.

\begin{figure*}[htbp]
  \centering
  \includegraphics[width= 0.48\linewidth]{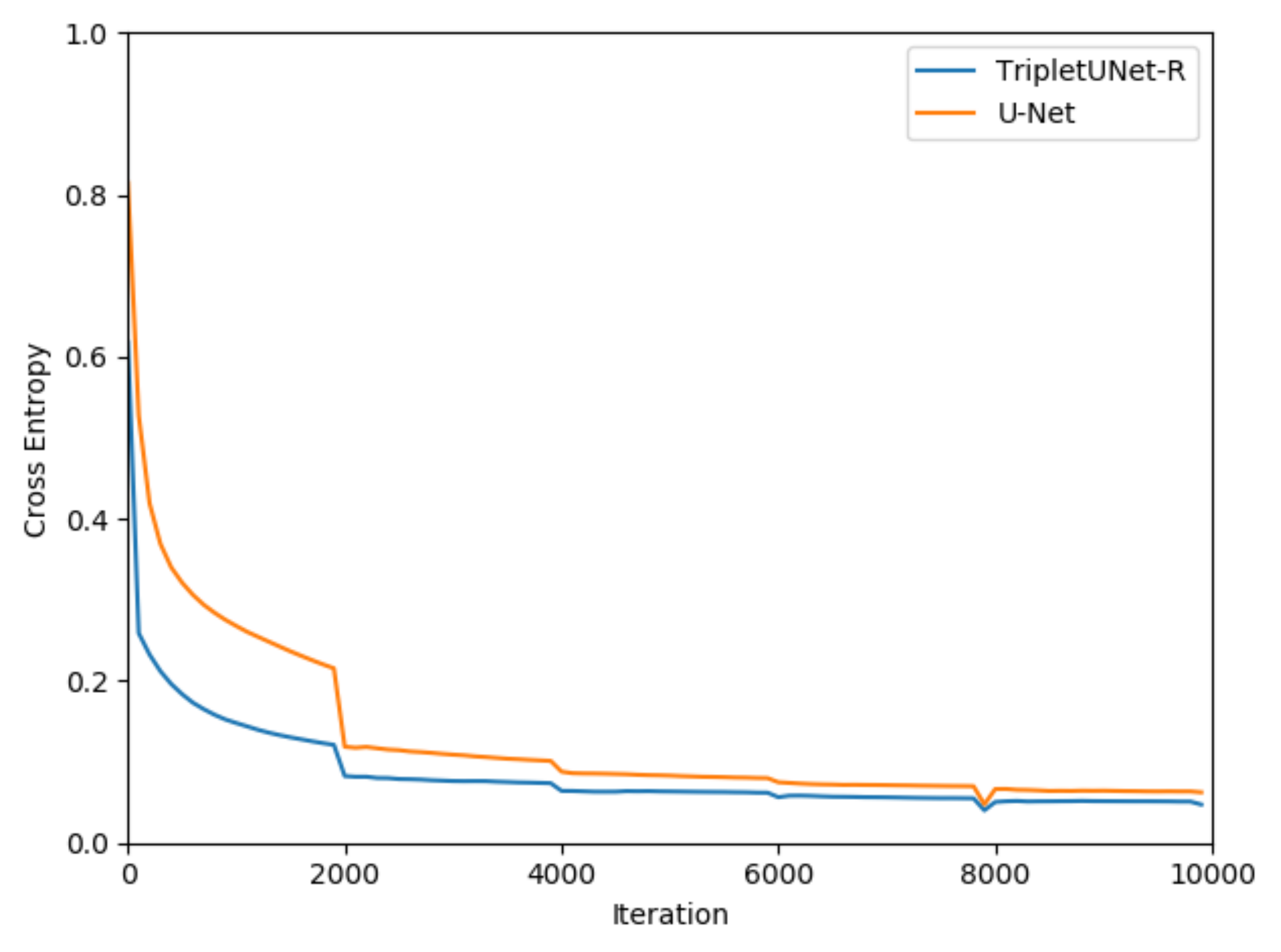} 
  \includegraphics[width= 0.48\linewidth]{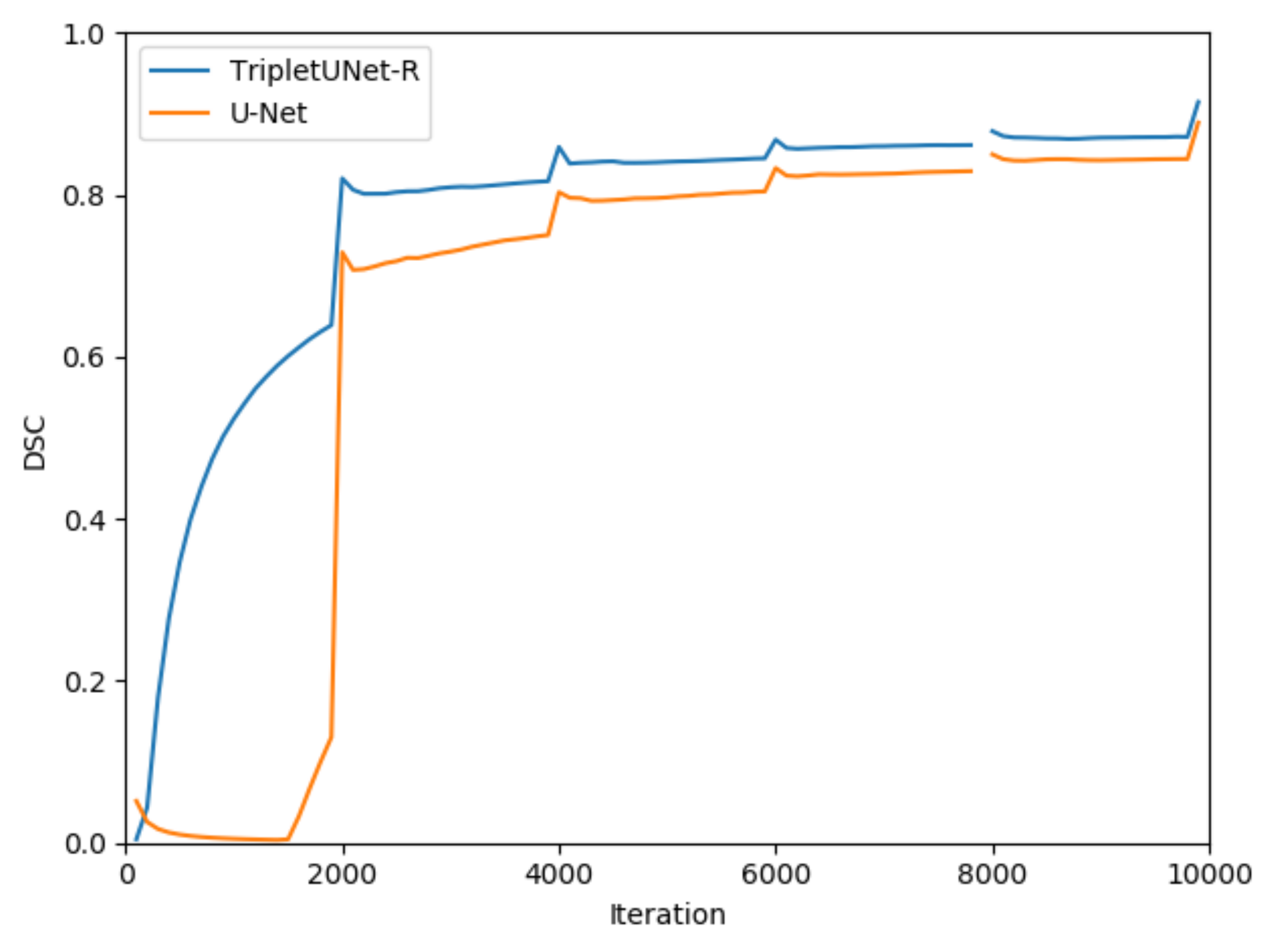}
  \caption{\label{fig:trainingloss} The convergence analysis of the proposed MetricUNet-R compared with UNet.}
\end{figure*}

The convergence analysis of the MetricUNet-R compared with the conventional UNet is illustrated in Fig. \ref{fig:trainingloss}. (a) and (b) in Fig. \ref{fig:trainingloss} show the cross-entropy loss and DSC curve of the two methods, respectively. The figures suggest that the incorporation of the proposed voxel-metric learning module can boost the convergence of the segmentation network, since the network obtained more guidance from the pixel features.

The comparison of segmentation performance among different MetricUNets are reported in Table \ref{Table:samplingstrategy}. It can be easily observed that, with the guidance of metric learning, the performance of the network is significantly improved. The performance in terms of DSC of the MetricUNets reported in the first bar in Table \ref{Table:samplingstrategy} (\ie, MetricUNet-R,H,C) is improved by $4.41\%$, $4.34\%$ and $4.46\%$, compared with baseline network UNet, which is trained by Cross-Entropy loss. Notably, in the proposed MetricUNets, we can observe that MetricUNet-R performs better among the three sampling methods. In DSC, MetricUNet-R method is slightly lower than MetricUNet-C; However, in ASD, MetricUNet-R outperform the MetricUNets with other two sampling methods by a margin. We conclude from the observation that random sampling can learn more integrated feature space in the proposed iteration based training scheme, since the other two sampling methods (\ie, focal hard negative sampling and contour-aware sampling) only sample from part of the whole feature space. Moreover, MetricUNet++-H with the UNet++ backbone improves the performance by $1.75\%$ in terms of DSC. The MetricUNet-HR and -HC networks, which are jointly learned on two sampling methods, both get a performance improvement on DSC and ASD, compared with MetricUNet-H and -C, which are only learned on one sampling methods. Moreover, as suggested by the table, MetricUNet-HR did not perform better than MetricUNet-HC, which reveals the combination of two sampling methods can have the chance to learn more integrated feature space. Thus the advantage of using random sampling is indistinctive.

\subsubsection{The influence of $\lambda$ and $\sigma$}

We compared the segmentation performance of MetricUNet-R in terms of different $\lambda$ and $\sigma$ to evaluate the influence of these two parameters. The networks were trained with the same settings of other hyper-parameters. We select several possible choices of hyper-parameters as limited by time and computational resources.
Fig. \ref{fig:comp_alpha} shows the performance in DSC and ASD of MetricUNet-R with $\lambda \in \{0.1,0.01,0.001\}$. The figure indicates that the performance of the proposed MetricUNet-R may vary according to the choice of $\sigma$. As shown in the figure, the network performed best w.r.t. $\lambda=0.01$.
Fig. \ref{fig:comp_alpha} also shows the performance in DSC and ASD of MetricUNet-R with $\sigma \in \{0.1,0.3,0.5,0.7,1.0\}$. It is suggested by the figure that the proposed network with $\sigma=0.7$ performs best over other compared MetricUNets. The network achieves higher mean DSC and lower mean ASD. Moreover, the outliers in ASD are also closer to the mean ASD with this network compared with others. 

\subsubsection{The influence of number of anchors $k$}

As a tuple-based learning method, we explore whether the number of anchors $k$ could affect the performance of the method. We compared the segmentation performance of MetricUNet-R in terms of different number of anchors. As limited by computational resources, we compare MetricUNet-R with a discrete number of anchors $k \in \{20,50,100,200\}$, and report the segmentation performance in Table \ref{Table:k}. As suggested by results, the performance variation of MetricUNet is small with the change of the number of anchors. Therefore, we use $k=20$ in this work, considering the efficiency of the network. In our experiments, MetricUNet-HP only takes less than $10\%$ additional training time compared with the conventional UNet.

\begin{table}[htbp]
\renewcommand{\arraystretch}{1.3}
\centering
\caption{\label{Table:k} Performance comparison of MetricUNet with respect to different number of anchors $k$.}
\setlength{\tabcolsep}{5pt}
\begin{tabular}{ccccc}
\toprule[1pt]
\textbf{$k$} & \textbf{DSC} & \textbf{ASD} & \textbf{SEN} & \textbf{PPV} \\
\toprule[1pt]
 20 & 0.88 $\pm$ 0.03 & 1.46 $\pm$ 0.77 & 0.87 $\pm$ 0.05 & 0.89 $\pm$ 0.04\\
50 & 0.88 $\pm$ 0.03 & 1.72 $\pm$ 0.77 & 0.89 $\pm$ 0.05 & 0.87 $\pm$ 0.07 \\
100 & 0.88 $\pm$ 0.02  & 1.51 $\pm$ 0.64 & 0.86 $\pm$ 0.05 & 0.89 $\pm$ 0.04  \\
200 & 0.88 $\pm$ 0.02  & 2.00 $\pm$ 1.37 & 0.88 $\pm$ 0.04 & 0.88 $\pm$ 0.04 \\
\toprule[1pt]
\end{tabular}\\
\end{table}

\subsection{Compare with the state-of-the-art methods}
The quantitative comparison of the proposed MetricUNet with the state-of-the-art methods is reported in Table \ref{Table:sota}. We compare MetricUNet with several remarkable methods including deep learning-based methods and deformable model-based methods:

\begin{itemize}
\item \cite{costa2007automatic} proposed a combination of deformable-model based method for the segmentation of prostate and bladder. The method is performed with different shape assumptions for the two organs.

\item \cite{shao2015locally} introduced a deformable model-based segmentation method for prostate and rectum in CT images, where a local boundary regression method is performed on the near-organ regions.

\item \cite{gao2016accurate} proposed a deformable model-based segmentation method combined with a random forest to obtain the organ boundary. The initialization problem of deformable models is thus alleviated.

\item \cite{dong2019synthetic} proposed a 3-D UNet based network with dilated convolution layers to improve the segmentation performance of prostate.

\item \cite{ronneberger2015u} proposed a fully convolutional network with two corresponding paths, namely UNet, with an encoding path and a decoding path with shortcut connections, so that the gradients in high-level can be better preserved to reach the low-level layers. The network is frequently used as the backbone and baseline in recent medical image segmentation studies.

\item \cite{milletari2016v} proposed the VNet, in which a Dice loss is used on a 3-D UNet based architecture with residual connections.

\item \cite{he2019pelvic} proposed a two-stage UNet based network to segment the pelvic organs. Specifically, a novel morphological representation, namely distinctive curve, is incorporated to provide additional guidance for the network.

\item \cite{nnUNet} proposed a general segmentation process for a bunch of medical image segmentation tasks. The method focuses on the choice of pre-processing techniques and hyper-parameters for specific datasets. And the network in this method is a slightly modified UNet.

\end{itemize}

The proposed method achieves the best overall segmentation performance among the listed methods. In terms of DSC, \cite{he2019pelvic}'s method outperform the other methods with a value of $0.87\pm0.03$. And \cite{nnUNet} achieves the best ASD results with a value of $1.61\pm1.05$. Compared with these method, our method obtains more than $1\%$ improvement in DSC, from $0.87$ to $0.88$; and $0.22mm$ improvement in ASD, from $1.61$ to $1.39$, which is a $15\%$ decreasing on the ASD value. The significant improvement on ASD indicates the effectiveness of the proposed method in delineating the organ contours, which is more valuable in clinical condition. Moreover, the proposed MetricUNet-HCP achieves $0.88$ and $0.89$ in mean SEN and PPV, respectively, which is better than the deformable model-based method in \cite{gao2016accurate} and the deep learning-based method in \cite{he2019pelvic,nnUNet}. 
By performing pair-wise t-test on the proposed method and the method in \cite{nnUNet}, the p-value indicates that the proposed method is significantly better than \cite{nnUNet}'s method in most metrics, \ie, DSC, SEN, ASD, and HD95. The tiny difference of the SEN and PPV value of the proposed method means that our proposed network is very robust at generating high quality segmentation, compared with the other methods. The robustness is also a key characteristic in clinical applications, where poor segmentation of the organ will lead to side effects. The conventional deep learning-based method in the second set of rows is not as good as the deformable models in the first set of rows. For example, VNet \cite{milletari2016v} achieves $0.85\pm0.04$ in DSC and $2.27\pm1.16$ in ASD, which is lower than the method in \cite{gao2016accurate}. This is because they were not specifically designed for the segmentation of prostate in CT images, which reveals the importance of incorporating domain knowledge and preserving the neighborhood information in the final segmentation map.

\subsection{Visualization Results}

\begin{figure}[htbp]
  \centering
  \includegraphics[width= 0.9\linewidth]{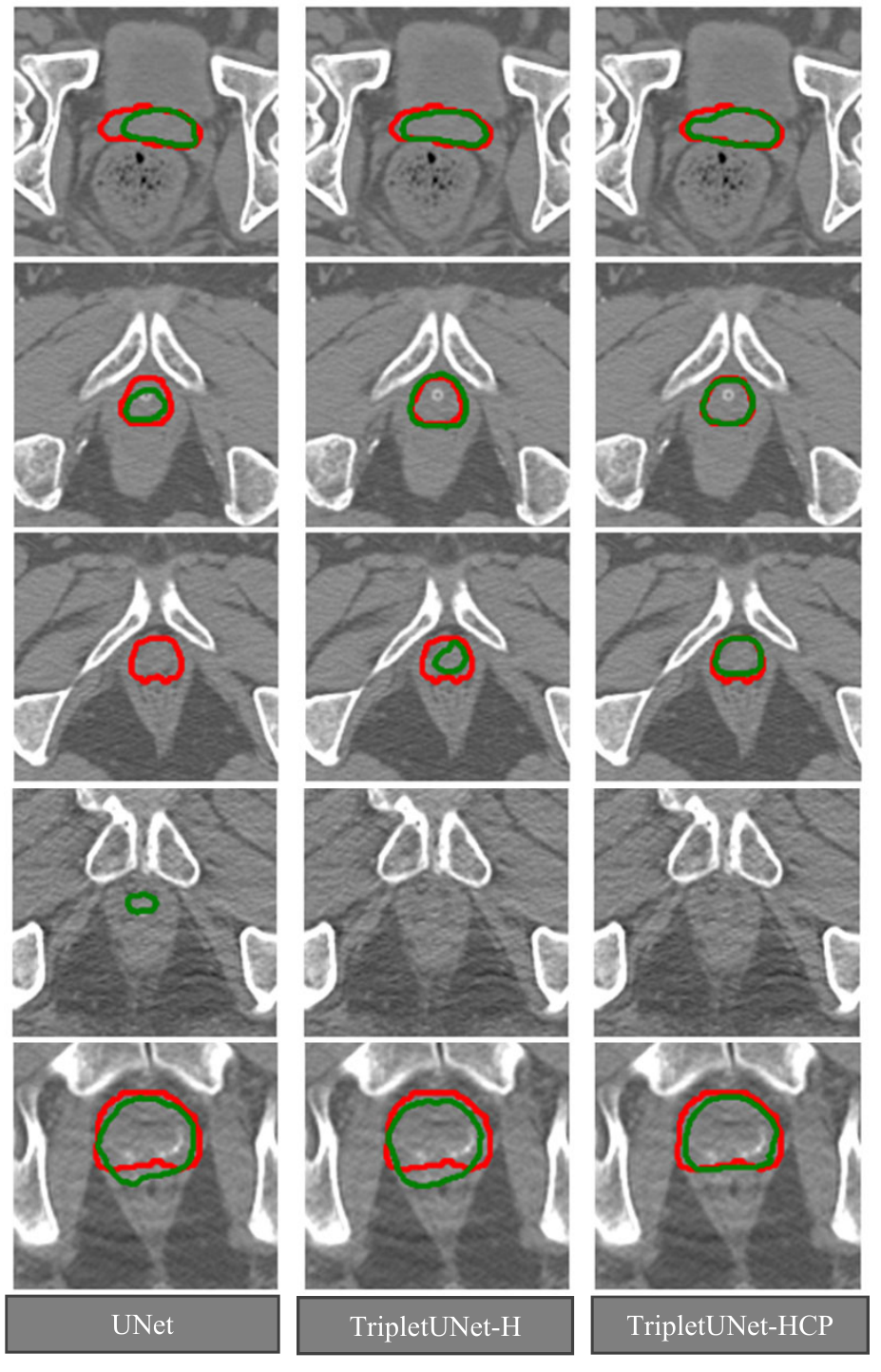}
  \caption{\label{fig:visres} The visualization of the results of UNet, MetricUNet-H and MetricUNet-HCP. Red contours denote the ground-truth, and green contours denote the generated segmentation masks.}
\end{figure}

We visualize the segmentation results of several typical cases generated by UNet, MetricUNet-H and MetricUNet-HCP in Fig. \ref{fig:visres}. As suggested by the figure, MetricUNet-HCP and MetricUNet-H can both generate more refined segmentations compared with the conventional UNet. Besides, MetricUNet-HCP performs better in some specific cases compared with MetricUNet-H. In very hard cases, over- or less-segmentation results are generated by UNet, while with voxel-wise metric learning, MetricUNets often avoid these problems.

\subsection{External Validation on PROMISE 2012 dataset}

To investigate the generalization ability of the proposed MetricUNet, we further perform MetricUNet-HCP on an MRI prostate dataset, \ie, PROMISE 2012. We partition the dataset randomly with $70\%$ of data as training, $10\%$ as validation, and $20\%$ as testing. We compare the proposed MetricUNet-HCP with several state-of-the-art methods. Specifically, \cite{wang2019deeply} proposed a method involving group-dilated convolutional layers for the accurate prostate segmentation. \cite{jia2019hd} proposed a network fused with a 3D segmentation backbone and a 2D boundary detector. The comparison of segmentation performance in terms of DSC, ASD, HD95, and aRVD is reported in Table \ref{Table:MRI}. As shown Table \ref{Table:MRI}, based on the conventional UNet, MetricUNet-HCP also achieved comparable results with several top-ranking methods.

\begin{table}[htbp]
\renewcommand{\arraystretch}{1.3}
\centering
\caption{\label{Table:MRI} Performance of MetricUNet compared with state-of-the-art methods on PROMISE 2012 validation dataset.}
\setlength{\tabcolsep}{5pt}
\begin{tabular}{ccccc}
\toprule[1pt]
\textbf{Methods} & \textbf{DSC} & \textbf{ASD} & \textbf{HD95} & \textbf{aRVD} \\
\toprule[1pt]
\cite{wang2019deeply} & 0.88 & \textbf{1.02} & 9.50 & 8.93 \\
\cite{nnUNet} & 0.88 & 1.24 & 3.95 & 6.45 \\
\cite{jia2019hd} & \textbf{0.91} & 1.36 & 3.93 & \textbf{5.10} \\
\hline
MetricUNet-HCP & 0.90  & 1.31  & \textbf{3.32} & 5.77 \\
\toprule[1pt]
\end{tabular}\\
\end{table}

\section{Conclusion}

In this paper, we introduced a two-stage framework to accurately segment prostate from raw CT images. Specifically, in the first stage, the region of the prostate is quickly localized by a lightweight network with down-sampled CT images. In the second stage, a multi-task UNet guided by both segmentation information and voxel-level feature relationship is proposed for generating the fine segmentation map of prostate. The voxel-level feature relationship is learned by the proposed online voxel-metric learning module. 

Despite the effectiveness of the proposed voxel-wise metric learning method, a more valuable contribution is that our voxel-metric learning method suggests a promising direction of applying metric learning techniques into FCN-based pixel-to-pixel or voxel-to-voxel predictions. We conducted extensive experiments on a planning CT prostate image dataset and a benchmark MRI prostate image dataset, showing that the quality of deep features learned by FCN can be further improved by formulating the problem into a pixel-level classification with metric learning paradigm, without including any additional learnable parameters. 
Benefit from the generated voxel-level tuples, we can easily bring metric-based semi-supervised learning and meta-learning methods into segmentation, which can be regarded as our future work.

Although the effectiveness of our method (\ie, MetricUNets) has been proved by the experiments, as a metric learning-based method, the computational efficiency of the proposed network should be further optimized. Naturally, how to boost the efficiency of metric learning method is still an open question. Another intuitive question is to define more suitable metric learning methods for FCNs. Theoretically, the proposed voxel-metric feature embedding module can be composed with every layer of the network, and thus reorder the features by learning with better designed metrics. Besides, designing appropriate sampling strategies for specific tasks is desired. For example, we can consider the displacement of organ contours in contour-aware sampling for the application of medical image segmentation. These topics define direction for future works.

\bibliographystyle{IEEEtran}
\bibliography{IEEEabrv,TripUNet}




\ifCLASSOPTIONcaptionsoff
  \newpage
\fi

\end{document}